\documentclass[12pt]{article} 

\usepackage{epsfig} 
\usepackage{amsfonts}
\usepackage{amssymb}
\usepackage{amsmath}  
\usepackage{amsbsy}  
\usepackage{wasysym}
 
\newlength{\largfig}
\def\beq{\begin{equation}} 
\def\eeq{\end{equation}} 
 
\def\beqn{\begin{eqnarray}} 
\def\eeqn{\end{eqnarray}}

\def\al{\alpha}

\def\wwjj{W^+W^-jj}
\def\zzlldec{e^+e^- \,\mu^+\mu^-}
\def\zzlndec{e^+e^- \,\nu_\mu \bar\nu_\mu}
\def\zzfl{\ell^+\ell^-\ell'^+\ell'^-}
\def\zztl{\ell^+\ell^-\nu\bar\nu}
\def\wwjj{W^+W^-jj}

\def\mc{\mathcal}
\def\mr{\mathrm}
\newcommand{\tw}{\textwidth}

\def\sla#1{\ifmmode% 
\setbox0=\hbox{$#1$}% 
\setbox1=\hbox to\wd0{\hss$/$\hss}\else% 
\setbox0=\hbox{#1}% 
\setbox1=\hbox to\wd0{\hss/\hss}\fi% 
#1\hskip-\wd0\box1 }

\newskip\humongous \humongous=0pt plus 1000pt minus 1000pt

\newif\ifdtup

\catcode`@=12 
 
\catcode`\@=11 
 
\@addtoreset{equation}{section} 

\def\theequation{\thesection.\arabic{equation}} 
 
\def\@normalsize{\@setsize\normalsize{15pt}\xiipt\@xiipt 
\abovedisplayskip 14pt plus3pt minus3pt% 
\belowdisplayskip \abovedisplayskip 
\abovedisplayshortskip \z@ plus3pt% 
\belowdisplayshortskip 7pt plus3.5pt minus0pt} 
 
\def\small{\@setsize\small{13.6pt}\xipt\@xipt 
\abovedisplayskip 13pt plus3pt minus3pt% 
\belowdisplayskip \abovedisplayskip 
\abovedisplayshortskip \z@ plus3pt% 
\belowdisplayshortskip 7pt plus3.5pt minus0pt 
\def\@listi{\parsep 4.5pt plus 2pt minus 1pt 
     \itemsep \parsep 
     \topsep 9pt plus 3pt minus 3pt}} 
 
\@twosidetrue

\catcode`\@=11  
\def\section{\@startsection{section}{1}{\z@}{3.5ex plus 1ex minus 
   .2ex}{2.3ex plus .2ex}{\large\bf}}

\def\thesection{\arabic{section}} 
\def\thesubsection{\arabic{section}.\arabic{subsection}} 
\def\thesubsubsection{\arabic{section}.\arabic{subsection}.\arabic{subsubsection}} 
 
\def\appendix{\setcounter{section}{0} 
 \def\thesection{\Alph{section}} 
 \def\theequation{\Alph{section}.\arabic{equation}} 
\def\thesubsection{\Alph{section}.\arabic{subsection}} 
\def\thesubsubsection{\Alph{section}.\arabic{subsection}.\arabic{subsubsection}} 
 
\def\section{\@startsection{section}{1}{\z@}{3.5ex plus 1ex minus 
   .2ex}{2.3ex plus .2ex}{\large\bf}} 
}

\newcommand{\ccaption}[2]{ 
  \begin{center} 
    \parbox{0.85\textwidth}{ 
      \caption[#1]{\small\it {#2}}} 
  \end{center}    } 
 
\def \to   {\mbox{$\rightarrow$}}

\newcount\minutes 
\newcount\scratch 
 
\def\timestamp{% 
\scratch=\time 
\divide\scratch by 60 
\edef\hours{\the\scratch} 
\multiply\scratch by 60 
\minutes=\time 
\advance\minutes by -\scratch 
---$\,$\hours:\null 
\ifnum\minutes< 10 0\fi 
\the\minutes}

\begin{document} 
\begin{titlepage} 
\nopagebreak 
{\flushright{ 
        \begin{minipage}{5cm} 
         Bicocca-FT-06-7 \\
         KA--TP--04--2006  \\       
         SFB/CPP--06--20 \\       
        {\tt hep-ph/0604200}\hfill \\ 
        \end{minipage}        } 
 
} 
\vfill 
\begin{center} 
{\LARGE \bf \sc 
 \baselineskip 0.9cm 
Next-to-leading order QCD corrections to $Z$ boson pair production via\\
vector-boson fusion 

} 
\vskip 0.5cm  
{\large   
Barbara J\"ager$^1$, Carlo Oleari$^2$ and Dieter Zeppenfeld$^1$ 
}   
\vskip .2cm  
{$^1$ {\it Institut f\"ur Theoretische Physik, 
        Universit\"at Karlsruhe, P.O.Box 6980, 76128 Karlsruhe, Germany}
}\\ 
{$^2$ {\it Dipartimento di Fisica "G. Occhialini", 
        Universit\`a di Milano-Bicocca, 
        20126 Milano, Italy}}\\   
 
 \vskip 
1.3cm     
\end{center} 
 
\nopagebreak 
%\vfill 
%\vskip 3cm 
\begin{abstract}
Vector-boson fusion processes are an important tool for the study of
electroweak symmetry breaking at hadron colliders, since they allow to 
distinguish a light Higgs boson scenario from
strong weak boson scattering. We here consider the channels 
$WW\,\to\, ZZ$ and $ZZ\,\to\, ZZ$ as 
part of electroweak $Z$ boson pair production in association 
with two tagging jets. 
We present the calculation of the NLO QCD corrections to the cross
sections for 
$pp\,\to \,\zzlldec + 2\; \mr{jets}$ and  $pp\,\to\,\zzlndec + 2 \; \mr{jets}$ 
via vector-boson fusion at order $\alpha_s\,\alpha^6$, which is
performed in the form a NLO parton-level Monte Carlo program.
The corrections to the integrated cross sections are found to be modest,
while the shapes of some kinematical distributions change appreciably at NLO.
Residual scale uncertainties typically are at the few percent level.

\end{abstract} 
\vfill 
%\today \timestamp \hfill 
\vfill 

% PACS: 14.80.Bn
\end{titlepage} 
\newpage               
%
%%%%%%%%%%%%%%%%%%%%%%%%%%%%%%%%%%%%%%%%%%%%%%%%%%%%%%%%%%%%%%%%%
%                                                       
\section{Introduction}
\label{sec:intro}
One of the primary goals of the CERN Large Hadron Collider (LHC) is the 
discovery of the Higgs boson and %with it 
a thorough investigation of
the mechanism of electroweak (EW) symmetry breaking~\cite{ATLAS,CMS}. 
In this context, vector-boson fusion (VBF) processes have emerged as a
particularly interesting  class of processes. 
Higgs boson production in VBF, i.e.\ 
the reaction $qq\,\to\, qqH$, where the Higgs decay products are detected in
association with two tagging jets, offers a promising
discovery channel~\cite{Rainwater:1999gg} and, once its existence
has been verified, will help to constrain the couplings of the Higgs
boson to gauge bosons and fermions~\cite{Zeppenfeld:2000td}. 

In order to distinguish possible signatures of strong weak-boson
scattering from those of a light Higgs boson, a good understanding of
$WW\,\to\, ZZ$  and $ZZ\,\to\, ZZ$ scattering processes, which are part of the
VBF reaction $qq\,\to\, qq ZZ$, is needed.    
This requires the computation of  next-to-leading order (NLO) QCD
corrections to the $qq\,\to\, qq\,ZZ$ cross section, including the leptonic
decays of the $Z$ bosons. Experimentally, very clean signatures are
expected from the $ZZ\,\to\, \zzfl$ decays in VBF with four charged leptons
in the final state, the disadvantage of this channel being a rather
small $Z\,\to\, e^+e^-$ or $Z\,\to\, \mu^+\mu^-$ branching ratio of about 3\%.
The $ZZ\,\to\, \zztl$ channel, with two undetected neutrinos, on the 
other hand, results in a larger number of events
due to the larger $Z\,\to\,\nu\bar\nu$  branching ratio~\cite{cahn:hzz}.

LO results for EW $ZZ\, jj$ production in VBF 
have been available for more than two decades. The first
calculations~\cite{zz_ewa} were 
performed employing the effective $W$ approximation~\cite{EWA},
where the vector bosons radiated off the scattering quarks are treated as
on-shell particles and, therefore, kinematical distributions
characterizing the tagging jets cannot be predicted reliably. In the
following years, exact calculations for $qq\,\to\, qq\,ZZ$ have been
completed, first without $Z$ boson decay~\cite{dic_zz}, and then
including leptonic decays of the $Z$ bosons within 
the narrow width approximation~\cite{Baur_zz}.

We go beyond these approximations and develop a fully-flexible parton
level Monte Carlo 
program, which allows for the calculation of cross sections and kinematical
distributions for EW $ZZ\,jj$ production via VBF at NLO QCD accuracy. 
The program is structured in complete analogy to the respective code for 
EW $W^+W^- \, jj$ production presented in Ref.~\cite{JOZ:WW}.
Here, we calculate the $t$-channel weak-boson exchange contributions to
the full matrix elements for
processes like $qq\,\to \,qq\,\zzlldec$ and $qq\,\to \,qq\,\zzlndec$ at 
$\mc{O}(\alpha^6\alpha_s)$. We consider all resonant and non-resonant
contributions giving rise to a 
four charged-lepton and a two charged-lepton plus two neutrino final state,
respectively. Contributions from weak-boson exchange in the $s$-channel
are strongly suppressed in the phase-space regions where VBF can be
observed experimentally and therefore disregarded throughout. We do not
specifically require the leptons and neutrinos to stem from a genuine
VBF-like production process, but also include diagrams where one or two
of the $Z$ bosons are emitted from either quark line. Diagrams, where
the final state leptons stem from a $\gamma\,\to\,\ell^+\ell^-$ decay or
non-resonant production modes, are also taken into account.
Finite-width effects are fully
considered. For simplicity, we nonetheless refer to the $qq\,\to \,qq\,\zzfl$ 
and $qq\,\to\,qq\,\zztl$ processes computed this way generically as 
``EW $ZZ\,jj$'' production.

The outline of the paper is as follows. In Sec.~\ref{sec:calc} we briefly
summarize the calculation of the LO and NLO matrix elements for EW $ZZ\,jj$
production making use of the helicity techniques of Ref.~\cite{HZ}.
Section~\ref{sec:res} deals with phenomenological applications of the
parton-level Monte Carlo program which we have developed. Conclusions are
given in Sec.~\ref{sec:concl}.
%
%%%%%%%%%%%%%%%%%%%%%%%%%%%%%%%%%%%%%%%%%%%%%%%%%%%%%%%%%%%%%%%%%
%                                                       
%
\section{Elements of the calculation}
\label{sec:calc}
The calculation of NLO QCD corrections to EW $ZZ\,jj$ production closely
resembles our earlier work for EW $W^+W^-$ production in association
with two jets~\cite{JOZ:WW}.
The main differences lie in the electroweak aspects of the processes, 
while the QCD structure of the NLO corrections is very similar. The
techniques developed in Ref.~\cite{JOZ:WW} can therefore be adapted
readily and only need a brief recollection here. 
For simplicity, we focus on the $\zzlldec$ decay channel in the following. 
The application of the basic features discussed for this case to the 
$\zzlndec$ leptonic final state is then straightforward.
%requiring the evaluation of a different set of Feynman diagrams only. 

\begin{figure}[!htb] 
\begin{center}
        \setlength{\unitlength}{1cm}
        \begin{picture}(17,19)
        \put(0.5,15){\includegraphics[width=0.35\tw,clip=]{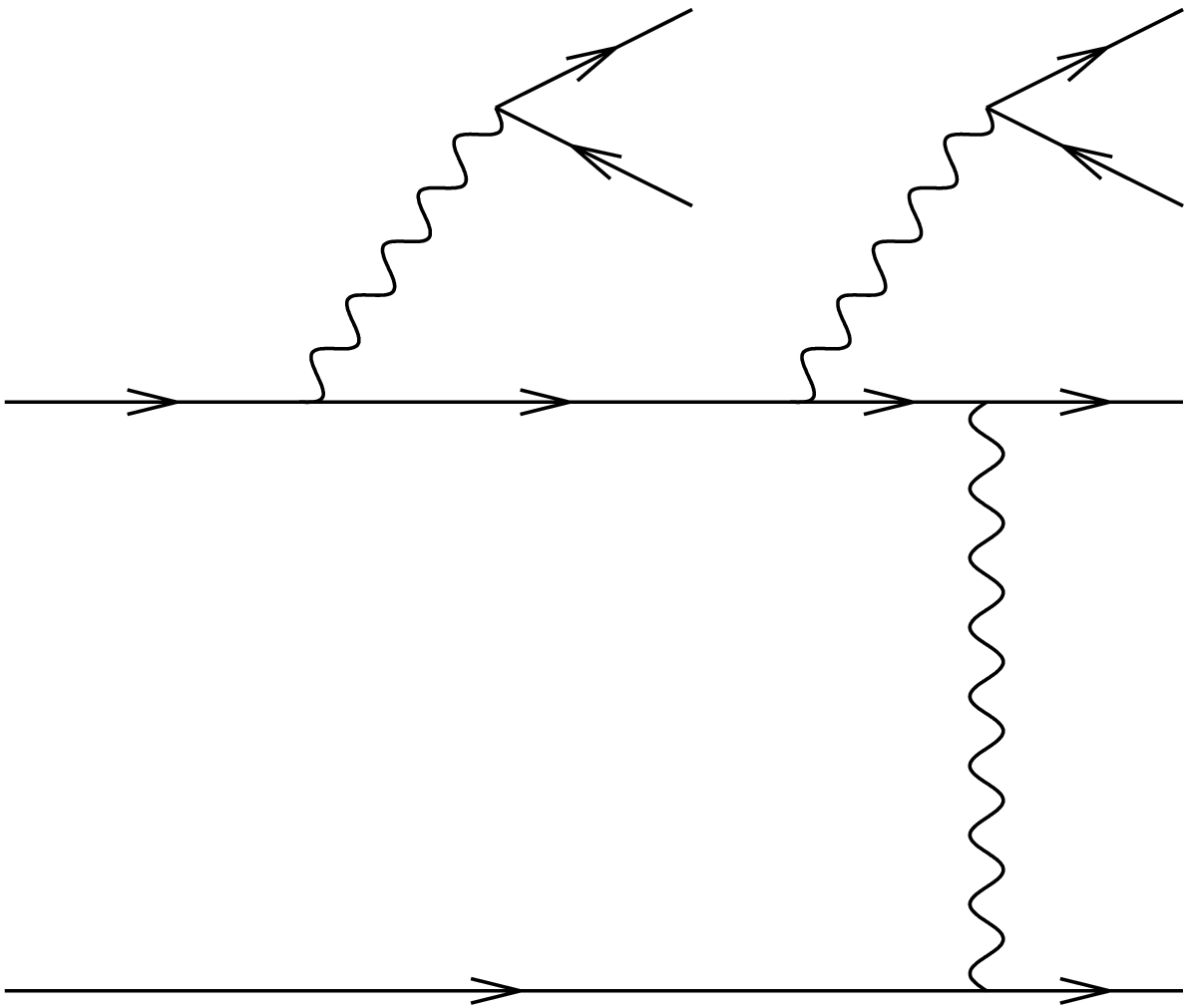}}
        \put(9.5,14.2){\includegraphics[width=0.35\tw,clip=]{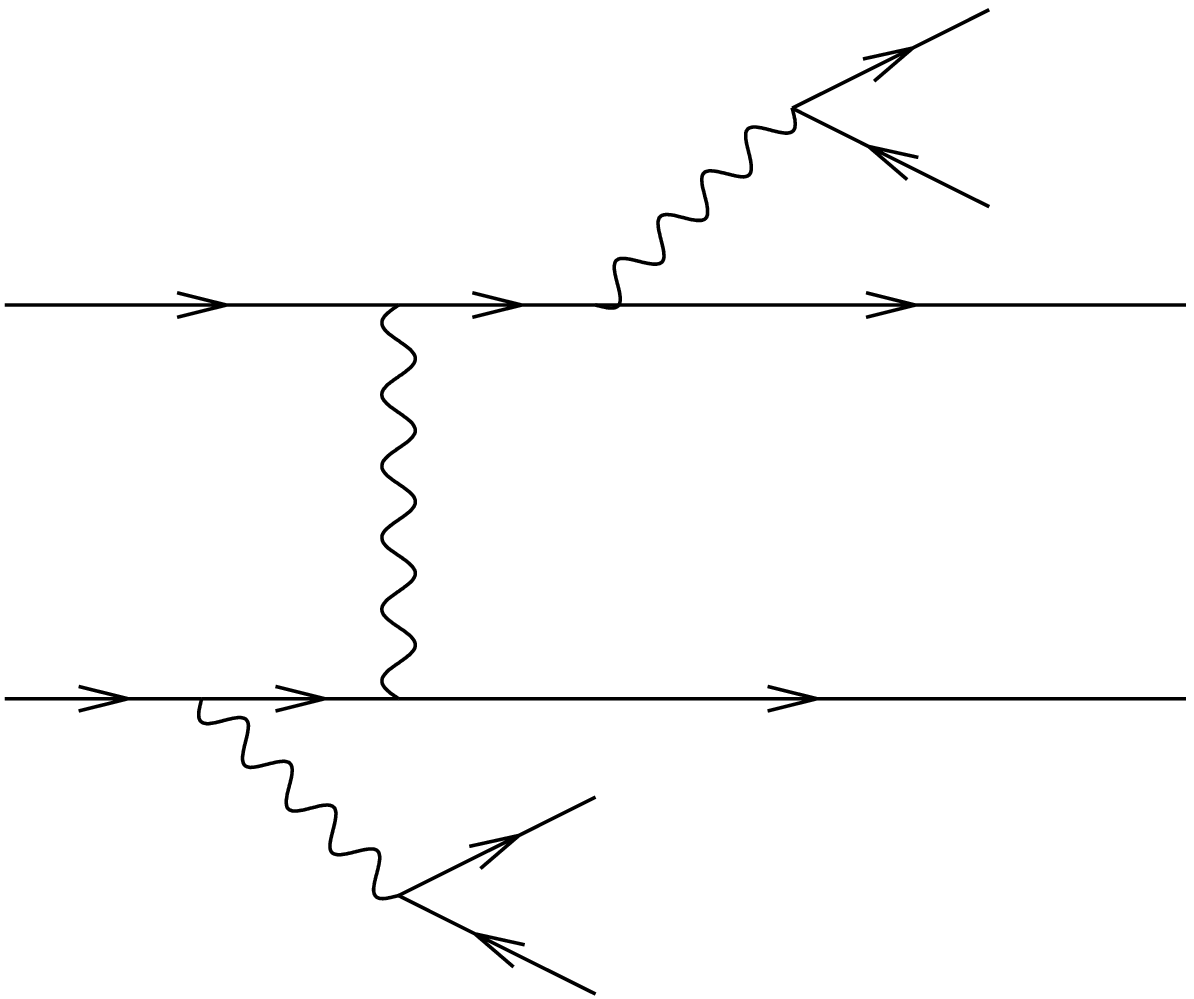}}
        \put(0.5,7.5){\includegraphics[width=0.35\tw,clip=]{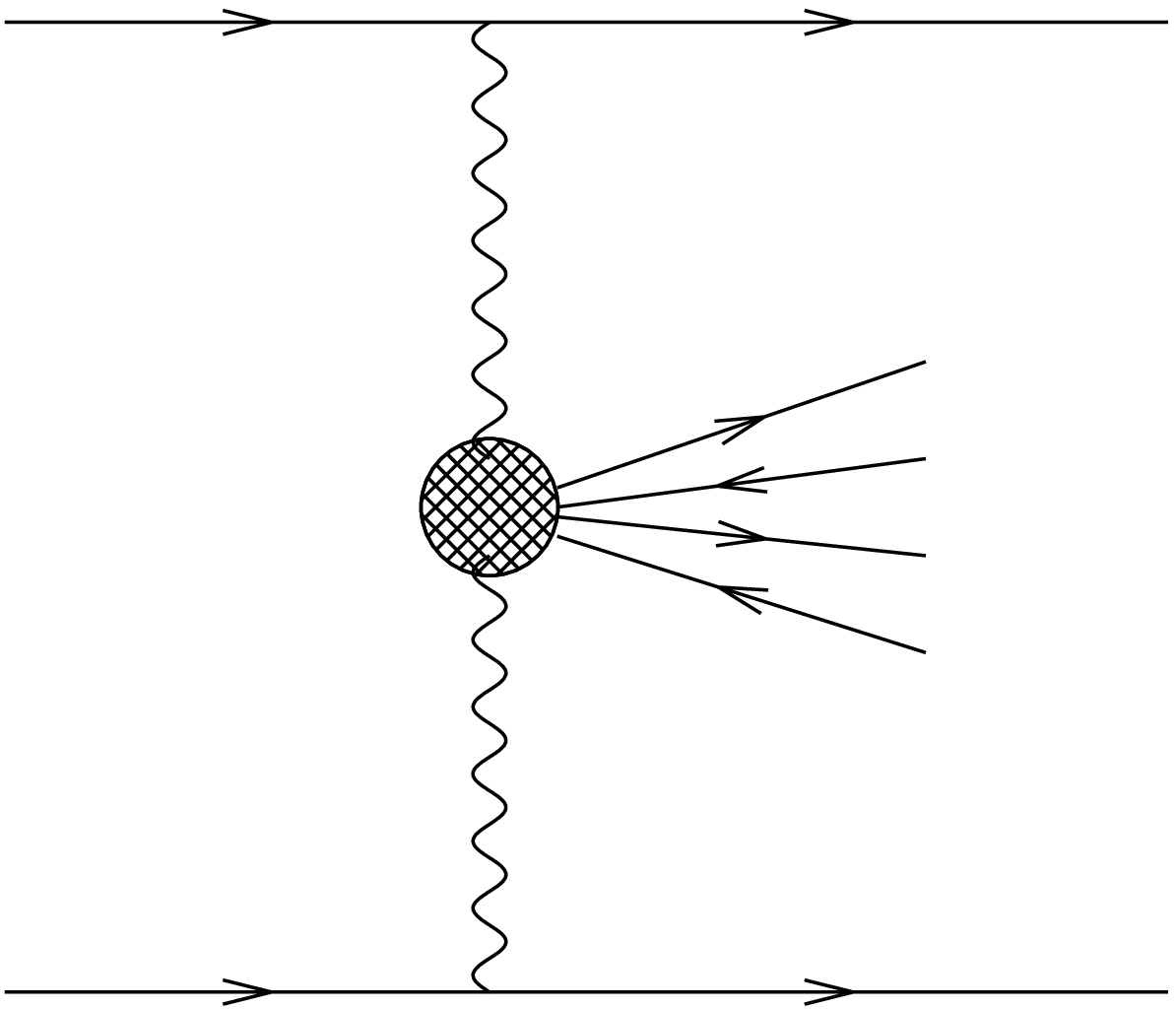}}
        \put(9.5,7.5){\includegraphics[width=0.35\tw,clip=]{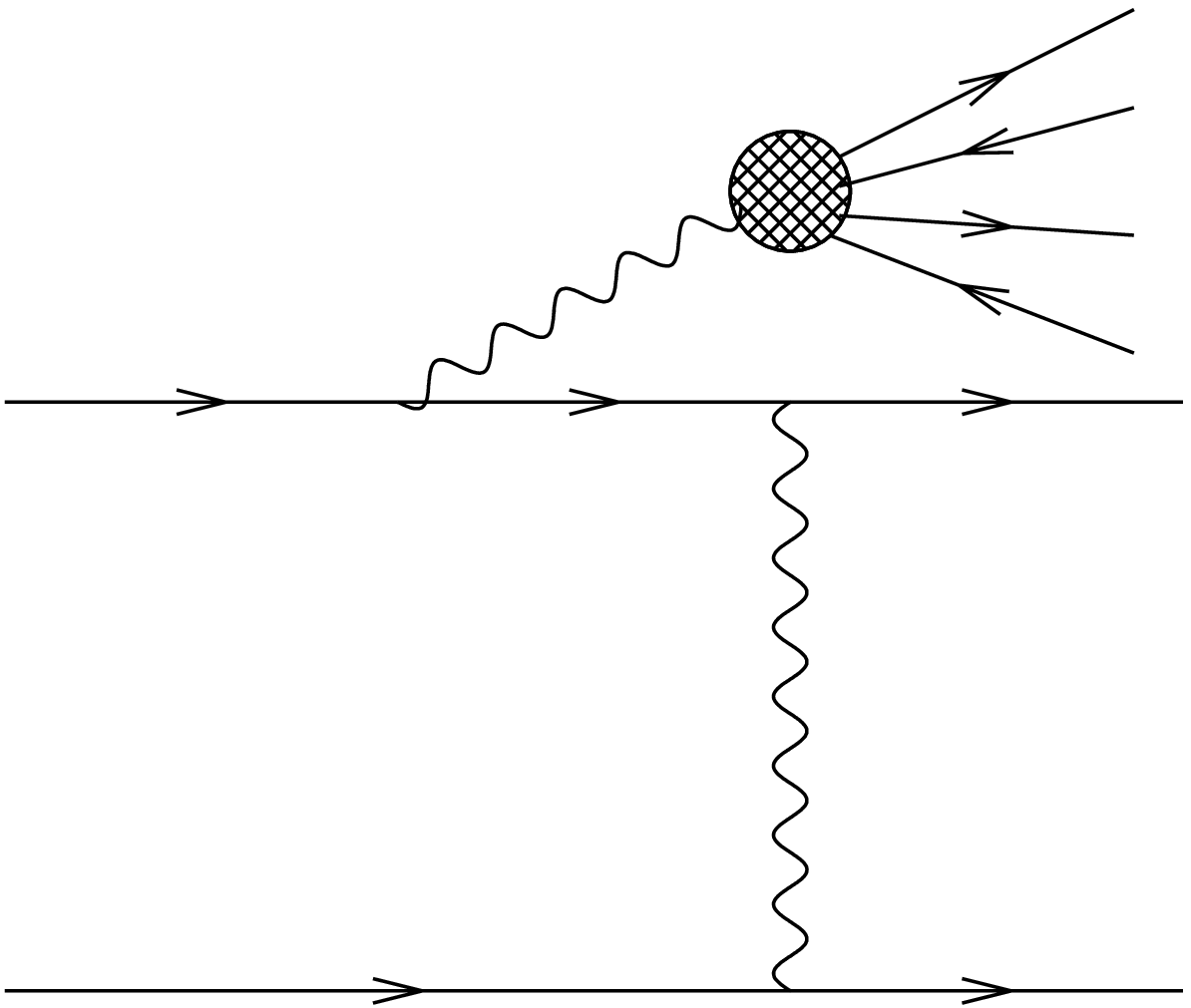}}
        \put(0.5,0.6){\includegraphics[width=0.35\tw,clip=]{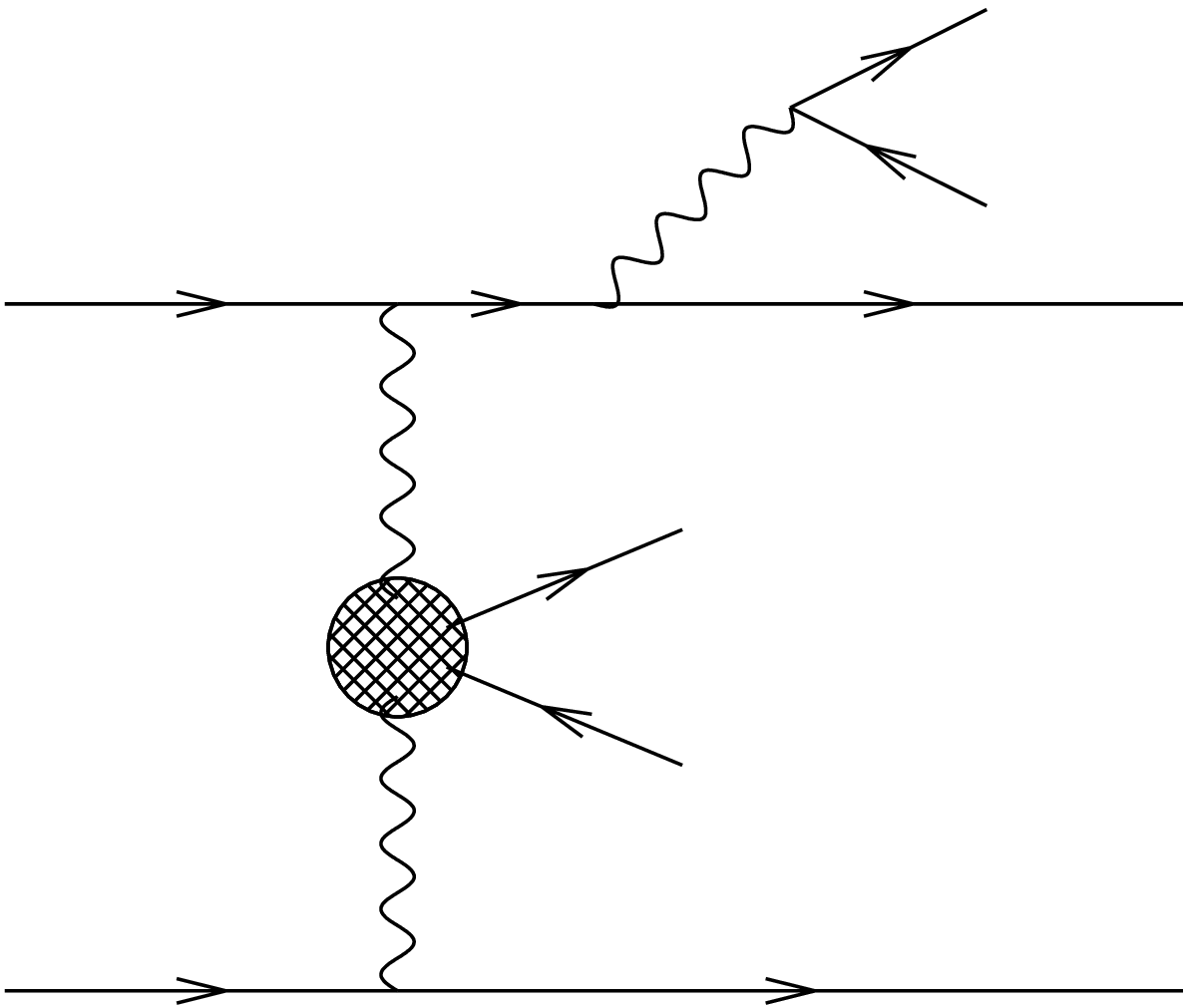}}
        \put(9.5,0.6){\includegraphics[width=0.35\tw,clip=]{zz-bv.eps}}
{\large 
        \put(0,17.8){$u$}
        \put(6.5,17.8){$u$}
        \put(0,15){$c$}
        \put(6.5,15){$c$}
        \put(4.5,16.5){$V$}
        \put(5.1,18.3){$V$}
        \put(2.7,18.3){$V$}
        \put(4,18.8){$e^+$}
        \put(4,19.8){$e^-$}
        \put(6.4,18.8){$\mu^+$}
        \put(6.4,19.8){$\mu^-$}
        \put(9,17.5){$u$}
        \put(15.5,17.5){$u$}
        \put(9,15.5){$c$}
        \put(15.5,15.5){$c$}
        \put(11.8,16.4){$V$}
        \put(12,17.9){$V$}
        \put(10.4,14.9){$V$}
        \put(14.6,17.9){$e^+$}
        \put(14.6,19.){$e^-$}
        \put(12.6,14.2){$\mu^+$}
        \put(12.6,15.2){$\mu^-$}
        \put(0,12.2){$u$}
        \put(6.5,12.2){$u$}
        \put(0,7.4){$c$}
        \put(6.5,7.4){$c$}
        \put(2.2,10.9){$V$}
        \put(2.2,8.3){$V$}
        \put(1.5,9.6){$L_{VV}^{\alpha\beta}$}
        \put(5.3,9.){$e^+$}
        \put(5.5,9.5){$e^-$}
        \put(5.5,10.1){$\mu^+$}
        \put(5.3,10.7){$\mu^-$}
        \put(9,10.3){$u$}
        \put(15.5,10.3){$u$}
        \put(9,7.4){$c$}
        \put(15.5,7.4){$c$}
        \put(12.7,8.9){$V$}
        \put(11.7,11.){$V$}
        \put(12.9,12.1){$\Gamma^\alpha_V$}
        \put(15.2,10.7){$e^+$}
        \put(15.2,11.2){$e^-$}
        \put(15.2,11.8){$\mu^+$}
        \put(15.1,12.4){$\mu^-$}
        \put(0,3.9){$u$}
        \put(6.5,3.9){$u$}
        \put(0,0.6){$c$}
        \put(6.5,0.6){$c$}
        \put(1.8,1.3){$V$}
        \put(1.8,3.2){$V$}
        \put(3.2,4.6){$V$}
        \put(0.8,2.1){$T_{VV,\mu}^{\alpha\beta}$}
        \put(5.4,4.5){$e^+$}
        \put(5.4,5.4){$e^-$}
        \put(4,1.7){$\mu^+$}
        \put(4,2.8){$\mu^-$}
        \put(9,3.9){$u$}
        \put(15.5,3.9){$u$}
        \put(9,0.6){$c$}
        \put(15.5,0.6){$c$}
        \put(10.8,1.3){$V$}
        \put(10.8,3.2){$V$}
        \put(12.2,4.6){$V$}
        \put(9.8,2.1){$T_{VV,e}^{\alpha\beta}$}
        \put(14.4,4.5){$\mu^+$}
        \put(14.4,5.4){$\mu^-$}
        \put(13,1.9){$e^+$}
        \put(13,2.8){$e^-$}
}       
        \put(3,13.4){(a)}
        \put(12,13.4){(b)}
        \put(3,6.8){(c)}
        \put(12,6.8){(d)}
        \put(3,0){(e)}
        \put(12,0){(f)}
        \end{picture}
\end{center}
\ccaption{} 
{\label{fig:feynBorn} 
The six Feynman-graph topologies contributing to the Born process $uc\,\to\,
uc\,\zzlldec$. Diagrams analogous
to (a), (d), (e) and (f), with vector-boson emission off the lower quark
line, are not shown.
}
\end{figure}

The Feynman graphs contributing to $pp\,\to \,\zzlldec\,jj$ can be
grouped in six topologies, respectively, for the 579 $t$-channel 
neutral-current (NC) and the 241 charged-current (CC) exchange diagrams
which appear at tree level.  
These groups are sketched in Fig.~\ref{fig:feynBorn}
for the specific NC 
subprocess $uc\,\to \,uc\,\zzlldec$. The first two of these correspond to the
emission of two external vector bosons $V$ from the same~(a) or different~(b)
quark lines. The remaining topologies are characterized  
by the vector-boson sub-amplitudes $L^{\al\beta}_{VV}$, $\Gamma^\al_V$,  
$T^{\al\beta}_{VV,\mu}$ and $T^{\al\beta}_{VV,e}$, 
which describe the tree-level 
amplitudes for the processes $VV\,\to\,\zzlldec$, $V\,\to\,\zzlldec$, 
$VV\,\to \,\mu^+ \mu^-$ and $VV\,\to\,e^+e^-$. 
In each case, $V$ stands for a virtual $\gamma$ or $Z$ boson, and 
$\al$ and $\beta$ are the tensor indices carried by these vector bosons.
The propagator factors $1/(q^2-m_V^2+im_V\Gamma_V)$ are included in the
definitions of the sub-amplitudes, which we 
call ``leptonic tensors'' in the following.
Graphs for CC processes such as $us\,\to \,dc\,\zzlldec$ are obtained by
replacing 
the $t$-channel $\gamma$ or $Z$ bosons in Fig.~\ref{fig:feynBorn} 
with $W$ bosons. They give rise to the new lepton tensors  
$L^{\al\beta}_{W^+W^-}$, 
$T^{\al\beta}_{W^+W^-,e}$ and $T^{\al\beta}_{W^+W^-,\mu}$ for the
sub-amplitudes $W^+W^-\,\to\,\zzlldec$, $W^+W^-\,\to \,e^+ e^-$ and 
$W^+W^-\,\to\,\mu^+\mu^-$.

Contributions from anti-quark initiated $t$-channel processes such as 
$\bar uc\,\to \,\bar uc\,\zzlldec$, which emerge from crossing the above 
processes, are fully taken into account. 
On the other hand, $s$-channel exchange
diagrams, where all vector bosons are time-like, contain vector-boson
production with subsequent decay of one of the bosons into a pair of
jets. These contributions can be safely neglected in the phase-space
region where VBF can be observed experimentally, with widely-separated
quark jets of large invariant mass. In the same way, $u$-channel
exchange diagrams are obtained by the interchange of identical 
final-state (anti)quarks. Their interference with the $t$-channel
diagrams is strongly suppressed for typical VBF cuts and therefore
completely neglected in our calculation.  

For the treatment of finite-width effects in massive vector-boson propagators
we resort to a modified version of the complex-mass 
scheme~\cite{Denner:1999gp}, which has already
been employed in Refs.~\cite{Oleari:2003tc,JOZ:WW}. We
globally replace vector-boson masses $m_V^2$ with $m_V^2-i m_V\Gamma_V$,
without changing the real value of $\sin^2\theta_W$. This procedure
respects electromagnetic gauge invariance.
The amplitudes for all NC and CC subprocesses are calculated and squared 
separately for each combination of external quark and lepton helicities. 
To save computer time, only the summation over the 
various quark helicities is done explicitely, while the four distinct 
lepton helicity states are considered by means of a random summation
procedure.

The computation of NLO corrections is performed in complete analogy to
Ref.~\cite{JOZ:WW}. For the real-emission contributions we encounter 
2892 diagrams for
the NC and 1236 for the CC processes, which are evaluated using the amplitude
techniques of Ref.~\cite{HZ} and the leptonic tensors introduced above.  
Singularities in the soft and collinear regions of phase space are regularized
in the dimensional-reduction scheme~\cite{DR_citation} 
with space-time dimension $d=4-2\epsilon$. 
The cancellation of these divergences with the respective poles from the
virtual contributions is performed by introducing the 
counter terms of the dipole subtraction method~\cite{CS}. Since the
color and flavor structure of our processes are the same as for Higgs
boson production in VBF, the analytical form of subtraction terms and
finite collinear pieces is identical to the ones given in
Ref.~\cite{Figy:2003nv}.
The finite parts of the virtual contributions are evaluated by
Passarino-Veltman tensor reduction~\cite{Passarino:1978jh}, which is
implemented 
numerically. Here, the fast and stable computation of pentagon tensor
integrals is a major 
issue, which is tackled by making use of Ward identities and mapping a large
fraction of the pentagon diagrams onto box-type contributions 
with the methods developed in~\cite{JOZ:WW}. The residual pentagon
contributions amount only to about 
1~$\permil$ of the cross sections presented below.

The results obtained for the Born amplitude, the real emission and the 
virtual corrections have been tested extensively. 
For the tree-level amplitude, we have performed a comparison to the fully
automatically generated results provided by MadGraph~\cite{madgraph}, 
and we found agreement at the $10^{-13}$~level. In the same way, the
real emission contributions have been checked.  For the latter, also QCD
gauge invariance has been tested, which turned 
out to be fulfilled within the numerical accuracy of the program.  
The numerical stability of the finite parts of the pentagon contributions 
is monitored by checking numerically that they satisfy electroweak 
Ward identities with a relative error less than $\delta=1.0$. 
This criterion is violated by about 3\% of the generated events. 
The contributions from these phase-space points to the finite parts 
of the pentagon diagrams are disregarded and the remaining pentagon parts  
are corrected by a global factor for this loss. 
In Ref.~\cite{JOZ:WW} we found that this procedure gives a stable result
for the pentagon contributions when varying the accuracy parameter
$\delta$. 
%
%%%%%%%%%%%%%%%%%%%%%%%%%%%%%%%%%%%%%%%%%%%%%%%%%%%%%%%%%%%%%%%%%
%                                                       
\section{Numerical results}
\label{sec:res}
The cross-section contributions discussed in the previous section are 
implemented in a fully-flexible parton-level Monte Carlo program for 
EW $ZZjj$ production at NLO QCD accuracy, very similar to the programs for 
$Hjj$, $Vjj$ and $\wwjj$ production in VBF described in 
Refs.~\cite{Figy:2003nv}, \cite{Oleari:2003tc} and \cite{JOZ:WW}. 

Throughout our calculation, fermion masses are set to zero and external 
$b$- and $t$-quark contributions are neglected. However, the code does
allow the inclusion of $b$-quark initiated sub-processes for NC exchange
where internal top-quark propagators do not occur.
For the Cabibbo-Kobayashi-Maskawa matrix,
$V_{CKM}$, a diagonal form equal to the identity matrix has been used, 
which yields the same results as a calculation employing the exact $V_{CKM}$
when the summation over all quark flavors is performed. 

We use the CTEQ6M parton distributions with
$\alpha_s(m_Z)=0.118$ at NLO, and the CTEQ6L1 set at LO~\cite{cteq6}. We chose
$m_Z=91.188$~GeV, $m_W=80.419$~GeV and 
$G_F=1.166\times 10^{-5}/$GeV$^2$ as electroweak input parameters. Thereof,
$\alpha_{QED}=1/132.54$ and $\sin^2\theta_W=0.22217$ are computed via LO
electroweak relations. To reconstruct jets from final-state partons, the
$k_T$ algorithm~\cite{kToriginal,kTrunII} is used with resolution
parameter~$D=0.8$. 

Partonic cross sections are calculated for events with
at least two hard jets, which are required to have 
\beq
\label{eq:cutspj}
p_{Tj} \geq 20~{\rm GeV} \, , \qquad\qquad |y_j| \leq 4.5 \, .
\eeq 
Here $y_j$ denotes the rapidity of the (massive) jet momentum which is 
reconstructed as the four-vector sum of massless partons of 
pseudorapidity $|\eta|<5$. The two reconstructed jets of highest transverse 
momentum are called ``tagging jets''. At LO, they are the final-state
quarks which are characteristic of vector-boson fusion processes.  
Backgrounds to VBF are significantly reduced by requiring a large rapidity
separation of the two tagging jets. We therefore impose the cut 
\beq
\label{eq:cutsyjj}
\Delta y_{jj}=|y_{j_1}-y_{j_2}|>4\; .
\eeq
Furthermore, the tagging jets are imposed to reside in opposite detector
hemispheres,
\beq
\label{eq:cutsh}
y_{j_1} \cdot y_{j_2} < 0\, ,
\eeq
with an invariant mass 
\beq
\label{eq:cutsmjj}
M_{jj} > 600~{\rm GeV}\;.
\eeq

These cuts render the LO differential cross section for $ZZjj$ finite, since
they enforce finite scattering angles for the two quark jets. For
the NLO contributions, initial-state singularities, due to collinear 
$q\,\to\, gq$ and $g\,\to\, q\bar q$ splitting, are factorized into the
respective quark and gluon distribution functions of the proton. 
Additional divergences, stemming from the $t$-channel exchange of
low-virtuality  photons in real-emission diagrams, 
are avoided by imposing a cut on the virtuality of the photon,
$Q^2_{\gamma,\mathrm{min}}=4$~GeV$^2$. Events that do not pass this cut give
rise to a collinear $q\,\to\, q\gamma$ singularity, which is considered to be
part 
of the QCD corrections to $p\gamma\,\to\, ZZjj$ and not calculated here. 

In the discussion of the $pp\,\to\,\zzfl\,jj$ channel, we focus on the leptonic 
final state $\zzlldec$ throughout. Results for the four lepton final
state with any combination of electrons and/or muons 
(i.e.\ $e^+e^-\mu^+\mu^-$, $e^+e^-e^+e^-$ and $\mu^+\mu^-\mu^+\mu^-$) are
obtained by multiplying the respective numbers for $e^+e^-\mu^+\mu^-$ by a
factor of two. This procedure neglects very small Pauli-interference
effects for identical charged leptons, however. Similarly, the
$\zzlndec$ combination on which we concentrate, is related to an
arbitrary two lepton plus two neutrino  final state  in the case of
$pp\,\to\,\zztl\,jj$ production: Here, a factor of six is needed to take
into account all neutrino species and two families of charged leptons.
To ensure that the charged leptons are well observable, we impose the
lepton cuts 
\beqn
&& p_{T\ell} \geq 20~{\rm GeV} \,,\qquad |\eta_{\ell}| \leq 2.5  \,,\qquad 
\triangle R_{j\ell} \geq 0.4 \, , \nonumber \\
&& m_{\ell\ell} \geq 15~{\rm GeV} \,,\qquad  
\triangle R_{\ell\ell} \geq 0.2 \, ,
\label{eq:cutspl}
\eeqn
where $\triangle R_{j\ell}$ and $\triangle R_{\ell\ell}$ denote the
jet-lepton and lepton-lepton separation in the rapidity-azimuthal 
angle plane. In addition, the charged leptons are required to fall between
the rapidities of the two tagging jets
\beq
\label{eq:cutsyl}
y_{j,min}  < \eta_\ell < y_{j,max} \, .
\eeq

In order to compute the full cross section for EW~$ZZjj$ production,
contributions from the Higgs boson resonance, $qq\,\to\,Hqq\,\to\,ZZ qq$,
as well as from the $ZZ$ continuum have to be considered. 
A representative for the latter, which effectively starts at the $Z$-pair
threshold, is obtained (for a Higgs mass below the $Z$-pair threshold) by 
imposing a cut on the four-lepton invariant mass of 
\beq
\label{eq:offres}
M_{ZZ} = \sqrt{(p_{\ell^+}+p_{\ell^-}+p_{\ell'^+}+p_{\ell'^-})^2} > 
m_H+10\;{\rm GeV}\;,
\eeq
and correspondingly for the $\zztl$ final state.
Since the contribution from the Higgs boson resonance has already
been computed in Ref.~\cite{Figy:2003nv}, we focus on the $ZZ$ 
continuum in the following, if not stated otherwise, and assume  
$m_H=120$~GeV. 

\begin{figure}[!thb] 
\centerline{ 
\epsfig{figure=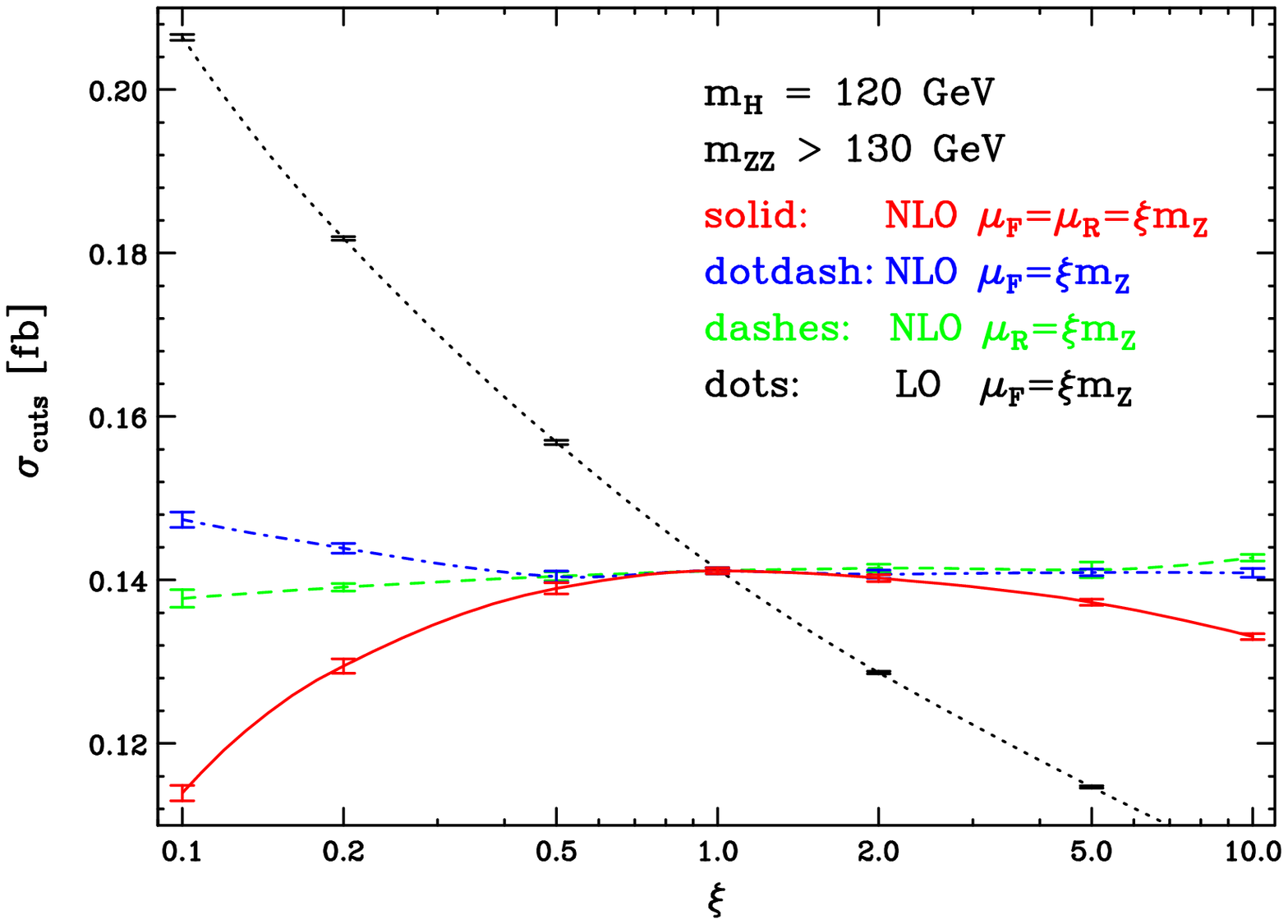,width=0.8\textwidth,clip=}
} 
\ccaption{} 
{\label{fig:scale_dep} 
Scale dependence of the total EW $\zzlndec\,jj$ cross
section at LO and NLO within the cuts of
Eqs.~(\ref{eq:cutspj})--(\ref{eq:offres}) for $pp$ collisions at 
the LHC. 
The NLO curves show $\sigma_{\rm cuts}^{\rm NLO}$ as functions of the scale
parameter $\xi$ for three different cases: $\mu_F=\mu_R=\xi m_Z$ (solid red), 
$\mu_F=\xi m_Z$ and $\mu_R=m_Z$ (dot-dashed blue), $\mu_F=m_Z$ and 
$\mu_R=\xi m_Z$ (dashed green). The LO cross section depends only on $\mu_F$
(dotted black line).
}
\end{figure}

The continuum cross section $\sigma_{\rm cuts}$ for EW $\zzlndec\,jj$
production,  
within the cuts of Eqs.~(\ref{eq:cutspj})--(\ref{eq:offres}), is shown in
Fig.~\ref{fig:scale_dep}. 
The figure illustrates the dependence of the NLO cross section on
the renormalization and factorization scales, $\mu_R$ and $\mu_F$, which are
taken as multiples of the $Z$ mass,
\beq
\label{eq:scale.mV}
\mu_R = \xi_R\,m_Z\;,\qquad\qquad \mu_F = \xi_F\,m_Z\; .
\eeq
The LO cross section only depends on $\mu_F= \xi_F\,m_Z$. By varying the scale
factor $\xi_F=\xi$ in the range $0.1 \div 10$, the value of
$\sigma_{\rm cuts}^{\rm LO}$ changes by almost a factor of two, indicating the
theoretical uncertainty of the LO calculation.
The strong scale dependence is reduced substantially after the inclusion
of NLO corrections. For $\sigma_{\rm cuts}^{\rm NLO}$ 
we study three different cases: $\xi_F=\xi_R=\xi$ (solid red line),  
$\xi_F=\xi$, $\xi_R=1$ (dot-dashed blue line), and $\xi_F=1$,
$\xi_R=\xi$ (dashed green line). The latter curve illustrates clearly
the weak dependence of  
$\sigma_{\rm cuts}^{\rm NLO}$ on the renormalization scale, which can be
understood from the fact that $\alpha_s(\mu_R)$ enters only at NLO,
while the LO cross section is completely independent of $\mu_R$. Also
the factorization-scale dependence of the full cross section is rather
low, such that the variation of $\sigma_{\rm cuts}^{\rm NLO}$ with the scale
parameter amounts to less than 2\% for all cases in the interesting
range $ 0.5<\xi<2$. In the following, we fix the scales to
$\mu_F=\mu_R=m_Z$.  

\begin{figure}[!thb] 
\centerline{ 
\epsfig{figure=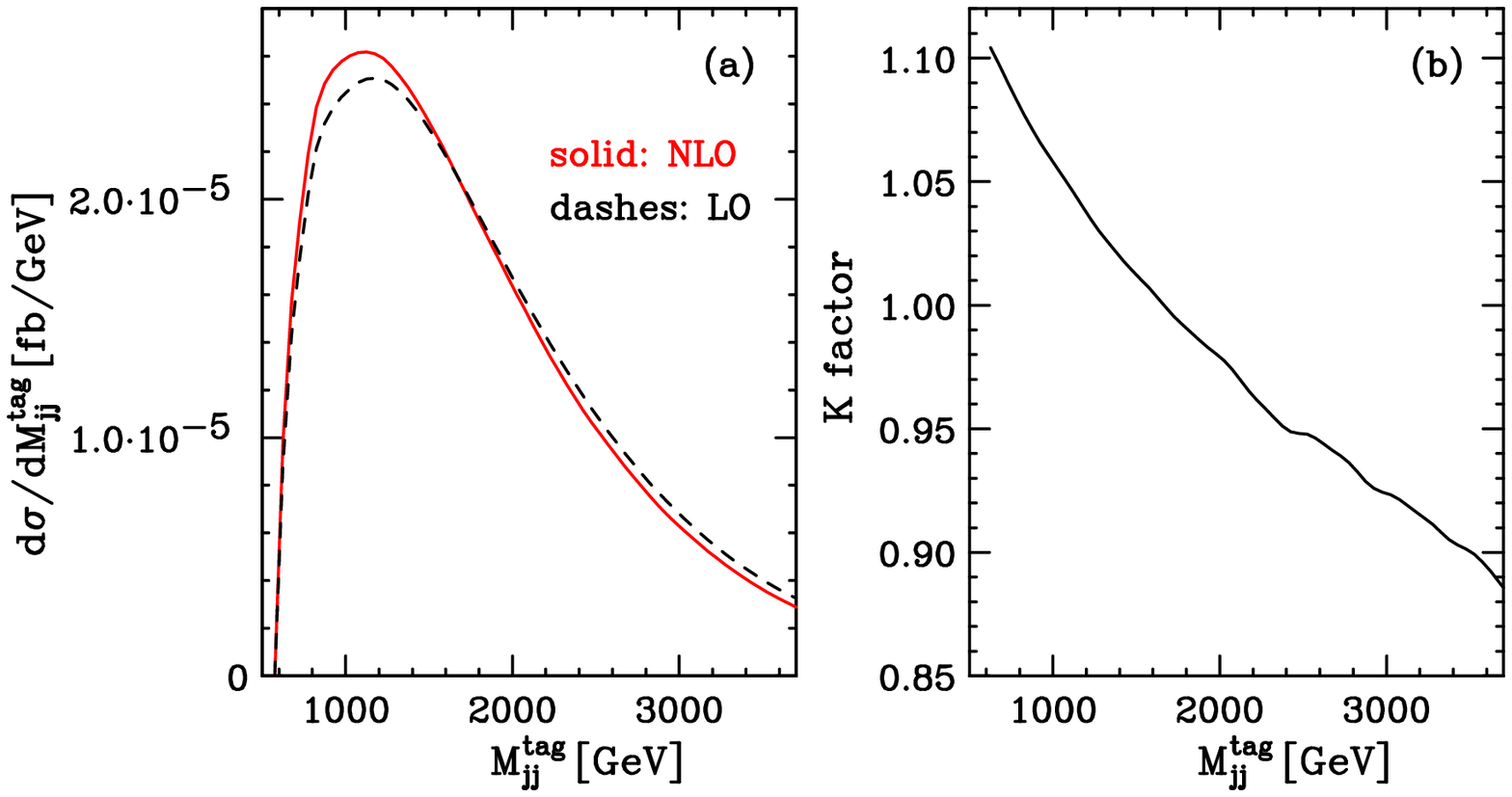,width=0.8\textwidth,clip=}
} 
\ccaption{} 
{\label{fig:mjj_max} 
Invariant-mass distribution of the tagging jets in EW
$\zzlldec\,jj$ production at the LHC. Panel~(a) shows the NLO (solid red)
and the LO results (dashed black). Panel~(b) displays the K factor as 
defined in Eq.~(\ref{eq:kfac}).
}
\end{figure}

As a representative for the observables characterizing the tagging jets, 
we show the invariant mass distribution of the tagging-jet pair,
$d\sigma/dM_{jj}^{\rm tag}$, for EW $\zzlldec\,jj$ production 
in Fig.~\ref{fig:mjj_max}.
The shape of the distribution at 
NLO differs from the respective LO result. This is emphasized in 
panel~(b) of the figure, where we show the dynamical K~factor, defined as
\beq
\label{eq:kfac}
K(x) = \frac{d\sigma_{\rm NLO}/dx}{d\sigma_{\rm LO}/dx}\;.
\eeq
Due to the extra parton emerging in the real-emission contributions to the NLO
cross section, the quarks which constitute the tagging jets tend to have 
smaller transverse momenta than at LO, thereby giving rise to 
lower values of their invariant mass.
The transverse-momentum distributions of the tagging jets, per se, exhibit  
K~factors in the range $0.8 \div 1.4$. On the other hand, 
angular distributions, such as the azimuthal angle between the
tagging jets,  
display rather uniform K~factors for the scale choice $\mu_F=\mu_R=m_Z$. 
For this particular scale choice, the total cross section is barely affected 
by the inclusion of NLO corrections, leading to a K~factor of 0.99.

\begin{figure}[!thb] 
\centerline{ 
\epsfig{figure=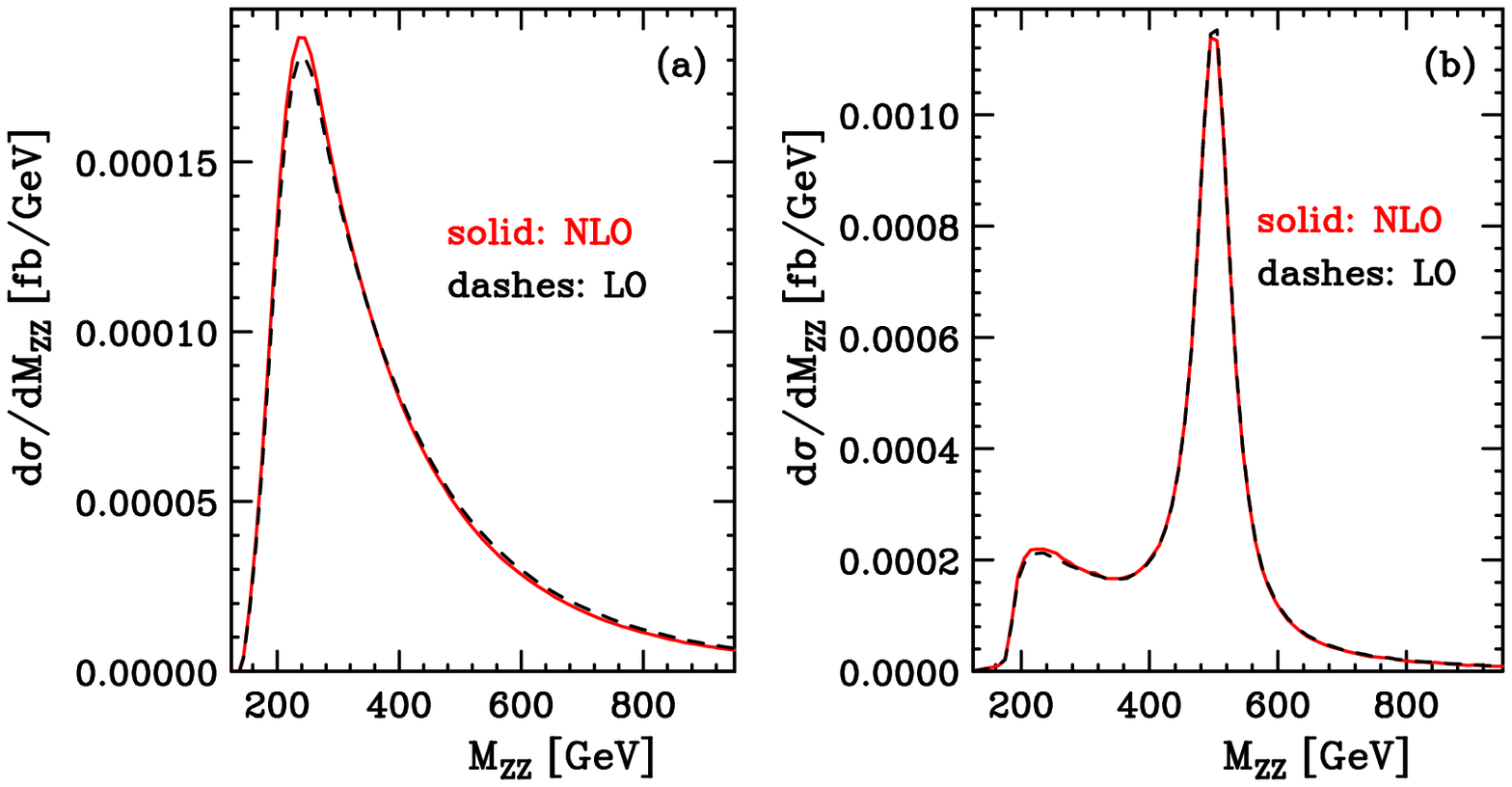,width=0.8\textwidth,clip=}
} 
\ccaption{} 
{\label{fig:mzz} 
Panel~(a): distribution of the four lepton invariant mass in EW $\zzlldec\,jj$ 
continuum production at the LHC, within the cuts of 
Eqs.~(\ref{eq:cutspj})--(\ref{eq:offres}) with $m_H=120$~GeV.
Panel~(b) shows the same observable when the contribution from
a Higgs boson of mass $m_H=500$~GeV is included.
In each case, NLO (red solid) and LO (black dashed) results are shown. 
}
\end{figure}

Finally, we show the distribution of the invariant mass of the $\zzlldec$
system, which is given by Eq.~(\ref{eq:offres})
and can be fully reconstructed experimentally. It is very sensitive to a light
Higgs boson, showing a pronounced resonance behavior for 
$m_H\lesssim 800$~GeV. 
For values of $m_H$ of the order of 1~TeV, the peak is diluted due to the
large corresponding width of the Higgs boson ($\Gamma_H \approx
500$~GeV) and the signal is 
distributed over a wide range in $M_{ZZ}$.
Figure~\ref{fig:mzz} 
illustrates the resonance behavior of $M_{ZZ}$: panel~(a) shows the $M_{ZZ}$ 
distribution for the 
continuum production of four charged leptons. Panel~(b) displays the same
observable, but now including the resonance contribution from a Higgs boson 
with $m_H=500$~GeV. 

One remarkable feature of Fig.~\ref{fig:mzz} is that
LO and NLO results are virtually indistinguishable, for the scale choice
$\mu_F=\mu_R=m_Z$. For the $M_{ZZ}$ distribution, the invariant mass of
the $Z$ pair is another possible scale choice, which, however, would lead to
substantially reduced LO differential cross section predictions at high
values of $M_{ZZ}$. One finds, for example, a decrease by a factor of 
$\approx 1.8$ % 1.77 +- 0.04
in $d\sigma^{\rm LO}/dM_{ZZ}$ at $M_{ZZ}=1.5\;{\rm TeV}$ when changing
$\mu_F=m_Z$ to $\mu_F=M_{ZZ}$, 
which largely is due to an underestimate of the LO parton
distributions at large Feynman-$x$.
The NLO prediction, on the other hand, decreases by about 13\% 
compared to 
our default choice $\mu_F=\mu_R=m_Z$, demonstrating the precision gained 
by including the NLO corrections.
%
%%%%%%%%%%%%%%%%%%%%%%%%%%%%%%%%%%%%%%%%%%%%%%%%%%%%%%%%%%%%%%%%%
%                                                       
\section{Conclusions}
\label{sec:concl}
In this paper we have presented first results for EW $ZZ\,jj$ production
at NLO QCD accuracy, obtained with a fully-flexible parton-level Monte
Carlo program that allows for a straightforward implementation of
realistic experimental cuts.  
The integrated cross sections for the two processes 
$pp\,\to\,\zzlldec\,jj$ and $pp\,\to \,\zzlndec\,jj$ 
were found to exhibit K~factors around 0.99, which shows that higher-order
corrections are under excellent control. Larger NLO
contributions are obtained for some kinematical distributions with dynamical
K~factors in the range $0.8\div 1.4$. These results hold for a default
scale choice of $\mu=m_Z$. Leading order results can change
substantially, by up to a factor of 2, for other scale choices, while NLO
results are very stable, demonstrating the value of the NLO corrections.
An estimate of the theoretical uncertainty of the NLO calculation
is provided by the scale variation of cross sections, within
VBF cuts. It amounts to about 2\% for integrated cross sections and, in
extreme cases, up to 10\% for distributions, when changing scales by a
factor of two. Similar uncertainties are induced
by the present errors on parton distribution functions, which, in the
analogous case of Higgs boson production in VBF, were found to be about
$\pm 3.5$\%~\cite{Figy:2003nv}.
%
%%%%%%%%%%%%%%%%%%%%%%%%%%%%%%%%%%%%%%%%%%%%%%%%%%%%%%%%%%%%%%%%%
%
\section*{Acknowledgements}
We are grateful to Gunnar Kl\"amke for useful discussions and to Stefan Gieseke
for support with our computer system.
This research was supported in part by the Deutsche Forschungsgemeinschaft
under SFB TR-9 ``Computergest\"utzte Theoretische
Teilchenphysik''. 
%
%%%%%%%%%%%%%%%%%%%%%%%%%%%%%%%%%%%%%%%%%%%%%%%%%%%%%%%%%%%%%%%%%
%

\end{document}